\documentclass{elsart}
\usepackage{amssymb}
\usepackage{epsfig,epsf,graphicx}
\usepackage{amsmath}
\def\be{\begin{eqnarray} &&}

\def\ee{\end{eqnarray}}

\def\bew{\begin{widetext}}
\def\ew{\end{widetext}}

\begin{document}
\begin{frontmatter}
\title{Scalar Mesons within a dynamical Holographic QCD model}

\author{W. de Paula and T. Frederico}
\address[ITA]{Departamento de F\'\i sica, Instituto Tecnol\'ogico de
Aeron\'autica, 12228-900 S\~ao Jos\'e dos Campos, SP, Brazil.}

\date{\today}
\maketitle
\begin{abstract}
We show that the infrared dynamics of string modes dual to $q\bar q$
states within a Dynamical AdS/QCD model of coupled dilaton-gravity
background gives the Regge-like spectrum of $f_0$'s scalars and
higher spin mesons consistent with experimental data. The pion mass
and its trajectory were also described with a scale deformation of
the metric and a rescaled string mass. The available experimental
decay widths of the $S\to PP$ decays provided a complementary check
of the proposed classification scheme for $f_0(980)$, $f_0(1370)$,
$f_0(1500)$ and $f_0(1710)$ as radial excitations of $f_0(600)$. For
$f_0(980)$ we estimated a mixing angle of $\pm 20^o$ with other
structures.
\newline\newline PACS{11.25.Tq, 11.25.Wx,14.40.-n,12.40.Yx}
\end{abstract}
\end{frontmatter}


The light scalar mesons are still challenging our imagination and
models for quark spectroscopy \cite{pdg}. Nowadays, tetraquarks
configurations\cite{Jaffe} are considered dominant for the structure
of $f_0(600)$ and $f_0(980)$ scalar resonances \cite{Maiani}. Also,
the decay of $f_0(1370)$ and $f_0(1500)$ into pions (2$\pi$ and
4$\pi$) suggest a structure of $n\bar{n}(=u\bar{u}+d\bar{d})$, while
the decay of $f_0(1710)$ into $K\bar{K}$ suggest a structure of
$s\bar{s}$\cite{pdg}. However those interpretations are not in
agreement with $\gamma\gamma$ colisions leading to
$K_{S}^{0}K_{S}^{0}$\cite{ACCIARRI01H} and $K^{+}K^{-}$\cite{ABE04}.
It leads to a possibility of interpret this resonances as a scalar
glueballs, or a mixing between those structures. The real nature of
scalars is a fundamental and controversial open question that needs
more experimental data to help to solve this issue.

In this work, using the Regge trajectories and scalar ($f_{0}$
family) decay widths into 2$\pi$ ($\Gamma_{\pi\pi}$) as guidelines,
we find that these states can still be interpreted as $q\bar q$
states in the framework of a Dynamical AdS/QCD model
\cite{dePaulaPRD09}. It has been applied with success to the
description of the Regge trajectories of light mesons with nonzero
spin. The model consists in the solution of five-dimensional gravity
(5d) coupled to an active dilaton field, where the metric comes from
a IR deformed anti-de Sitter space in order to have confinement by
the area law behavior of the Wilson loop. We expect that the
phenomenological 5d holographic QCD viewpoint
(\cite{PolchinskiPRL2002,Karch,BrodskyPRL09,Kiritsis,FBF,FK,dePaulaPRD09}),
consistent with experimental data, can give some hints to construct
a 10 dimensional (10d) dual model of QCD-like theories.

The AdS/CFT conjecture\cite{Maldacena_Conjecture} relates a 10d Type
IIB superstring field theory in $AdS_{5}\times S_{5}$ with
correlators of an N=4 super Yang-Mills theory in a four dimensional
flat space-time. The low energy limit of massless fields of a Type
IIB superstring theory is described by a Type IIB supergravity.
Therefore it is natural that the search for a QCD-like dual model
starts from a 10d solution of the Type IIB supergravity theory. In
this respect, an important step in this program is the investigation
of solutions of Type IIB supergravity that are pure N=1 super Yang
Mills in the IR. These solutions lead to dual models with some QCD
properties as confinement and chiral symmetry breaking. There are
some examples of supergravity solutions as Klebanov-Strassler
\cite{Klebanov:2000hb}, Klebanov-Tseytlin \cite{Klebanov:2000nc} and
Maldacena-Nunez \cite{Maldacena:2000yy} which have some
characteristics of an N=1 theory. On the other hand, the spectrum of
those models are still far from the phenomenology. An interesting
result of \cite{Berg:2006xy} is to show how to construct an
effective 5d action with several scalars starting with a 10d
consistent truncation of type IIB supergravity described by
Papadopoulos-Tseytlin (PT) ansatz\cite{Papadopoulos:2000gj}. In the
5d effective action, each scalar has a different sigma model metric.
It suggests that it may be conceivable a possibility that
phenomenological 5d effective models of QCD could play a role in
describing  mesons, and more generally hadrons.

Here we focus on the light-scalar and pseudoscalar sector of QCD.
Our results show that an appropriate choice of the scaling factor in
the 5d metric model gives the experimental Regge behavior (scalars
and pseudoscalars) and partial decay width of $f_0$'s into two
pions. The partial widths are obtained without introducing any free
parameter beyond those implicit contained in the description of the
scalar and pion Regge trajectories. We made the assumption that the
holographic coordinate in the metric of the Dynamical AdS/QCD model
for high spin mesons \cite{dePaulaPRD09} can be rescaled  in order
to provide the effective potential for the string modes dual to the
scalar and pseudoscalar mesons. The masses of the $f_0$ scalars are
then obtained by a single rescale of the holographic coordinate in
the metric form. We also describe the light pseudoscalar states. The
scale of the holographic coordinate is contracted for the
pseudoscalars in respect to the scalar case and a rescaled string
mass is also introduced to allow a description of the pion mass.

The outcome from our work is that $f_0(980)$, $f_0(1370)$, $f_0(1500)$, $f_0(1710)$, $f_0(2020)$, $f_0(2100)$, $f_0(2200)$ and $f_0(2330)$ are
excited states of $f_0(600)$, and they belong to a Regge trajectory with a slope of about 0.5 $GeV^2$  half of the corresponding one for the
$\rho$-meson trajectory.

Additionally to support our interpretation of the nature of the light $f_0$'s scalars, we obtain the wave functions of the string modes dual to
the scalar meson states and using the pion string amplitude we calculate the decay width of the scalars into two pions. The experimental
observation of the systematic decrease found for $\Gamma_{\pi\pi}$ for higher excitations is easily understood within the holographic view: the
overlap between the radial excitations and the two pion state amplitudes are depleted by increasing the number of nodes of the scalar wave
function. From that the decreasing pattern for the $S\to PP$ decay width appears.

The two-pion partial decay width for $f_0(600)$, $f_0(1370)$, $f_0(1500)$ and $f_0(1710)$, for which experimental information exists, are
qualitatively consistent with the model results. No further parameter is required besides the coupling between the scalar field with the pions
that is determined from the analysis of the pion mass.  In particular for $f_0(600)$ a width of about 600 MeV is found.

Although $f_0(980)$ has a mass identified with the first excitation of the string mode dual to $q\bar q$ state, it has a too large
$\Gamma_{\pi\pi}$ compared to the range of the experimental values. It is known that  $f_0(980)$ should mix strongly with other structures such as
$s \bar s$ states (see e.g. \cite{Bediaga}). Indeed we got a mixing angle of about $\pm 20^o$\cite{pdg} to be consistent with the experimental
data for the partial decay width. (We do not determine the sign of the mixing angle.) Below we briefly discuss the basis of the present
holographic model and substantiate quantitatively our claims.

The search for string dual models of Gauge theories was pursued
since the pioneering work by t' Hooft \cite{tHooft} and vigorously
developed after the Maldacena's Conjecture
\cite{Maldacena_Conjecture}. Applications of AdS/CFT to describe QCD
in a bottom-up approach started with the Hard Wall model
\cite{PolchinskiPRL2002} that provides a good description of form
factors at high $Q^{2}$, glueball mass spectrum\cite{Boschi}, but
does not give the linear Regge trajectory presented by the mesonic
data. This phenomenological result can be obtained by a spin
dependence in the metric \cite{FBF} or by introducing a dilaton
field\cite{Karch}(see also\cite{Afonin:2009pd}). In particular,
scalar mesons were analyzed in (\cite{Schmidt_Scalar} and
\cite{Colangelo_Scalar}). However, both works do not include the
sigma meson in their studies, while in \cite{kelley} the sigma is
associated to the ground state. Those models are not solutions of
Einstein equations and also do not confine by the Wilson Loop
criteria. We proposed a Dynamical AdS/QCD model \cite{dePaulaPRD09}
that is a solution of the Einstein equations, confines by the Wilson
loop criteria and describes the scalar sector. We start from the
Einstein-Hilbert action of five-dimensional gravity coupled to a
dilaton field $\Phi $,
\begin{equation}
S=\frac{1}{2\kappa ^{2}}\int d^{4}x dz\sqrt{\left\vert g\right\vert }\left( -%
\emph{R}+\frac{1}{2}g^{MN}\partial _{M}\Phi \partial _{N}\Phi -V(\Phi )\right) ,  \label{actiongd}
\end{equation}%
where $\kappa $ is the five-dimensional Newton constant, the dilaton field $\Phi \left( z\right) $ depends on the radial coordinate only and
$V(\Phi)$ is the potential for the dilaton field. We restrict our metric to $g_{MN}=e^{-2A(z)}\eta _{MN}$ with $\eta =$ diag$\left(
1,-1,-1,-1,-1\right) $.  Our model belongs to the general class of ``Improved AdS/QCD theories" proposed recently by G\"{u}rsoy, Kiritsis and
Nitti \cite{Kiritsis}.

The static solutions of the Einstein-dilaton coupled field equations in the fifth dimension found in \cite{dePaulaPRD09} are given by $\Phi
^{\prime }=\sqrt{3}\sqrt{A^{\prime }{}^{2}+A^{\prime \prime }}$ and $V(\Phi \left( z\right) )=3e^{2A\left( z\right)}\left[ A^{\prime \prime }
\left( z\right) -3A^{\prime }{}^{2}\left( z\right) \right]/2.$

The boundary condition on the physical brane restricts the geometry to asymptotically $\mathrm{AdS}_{5}$ ($\mathrm{ AAdS}_{5}$) space-times and
thus ensures conformality in the ultraviolet (UV).  The infrared (IR) behavior is chosen to obtain Regge-trajectories consistent with the
Wilson-loop area-law for the gravity dual of a confining theory. The conformal invariance is broken in the IR by $\Lambda_{QCD}$.

To calculate the spectrum of scalar mesons $\varphi$ we start from the action
\begin{equation}
I = \frac{1}{2} \int d^{4}x dz\sqrt{\left\vert g\right\vert } \left(g^{\mu\nu}\partial_{\mu}
\varphi(x,z)\partial_{\nu}\varphi(x,z)-\frac{M_{5}^{2}}{\Lambda_{QCD}^{2}}\varphi^{2} \right),
\end{equation}%
that describes a scalar mode propagating in the dilaton-gravity background. We factorize the holographic coordinate dependence as
$\varphi(x,z)=e^{iP_{\mu}x^{\mu}}\varphi(z)$, $P_{\mu}P^{\mu}=m^{2}$.

The string modes of the massive scalar field $\varphi$ can be rewritten in terms of the reduced amplitudes $\psi _{n}=\varphi_{n} \times
e^{-(3A+\Phi)/2}$ which satisfy the Sturm-Liouville equation
\begin{equation}
\left[ -\partial _{z}^{2}+\mathcal{V}(z)\right] \psi _{n}=m_{n}^{2}\psi _{n}  \label{sleq}
\end{equation}%
where the string-mode potential is
\begin{equation}
\mathcal{V}(z)=\frac{B^{\prime }{}^{2}(z)}{4}-\frac{B^{\prime \prime }(z)}{2}+M_{5}^{2}e^{-2A(z)},  \label{vs}
\end{equation}%
with $B=3A+\Phi$. (Note that $B=(2S-1)A+\Phi$ for the spin nonzero states \cite{Karch}.) The gauge/gravity dictionary identifies the eigenvalues
$m_{n,S}^{2}$ with the squared meson mass spectrum of the boundary gauge theory.

The AdS/CFT correspondence states that the wave function should behave as $z^{\tau}$, where $\tau = \Delta - \sigma$ (conformal dimension minus
spin) is the twist dimension for the corresponding interpolating operator that creates the given state configuration \cite{PolchinskiPRL2002}. The
five-dimensional mass chosen as \cite{Witten98} $M_{5}^{2}=\tau(\tau-4) \ ,$ fixes the UV limit of the dual string amplitude with the twist
dimension.

\begin{figure}[tbh]
\centerline{\epsfig{figure=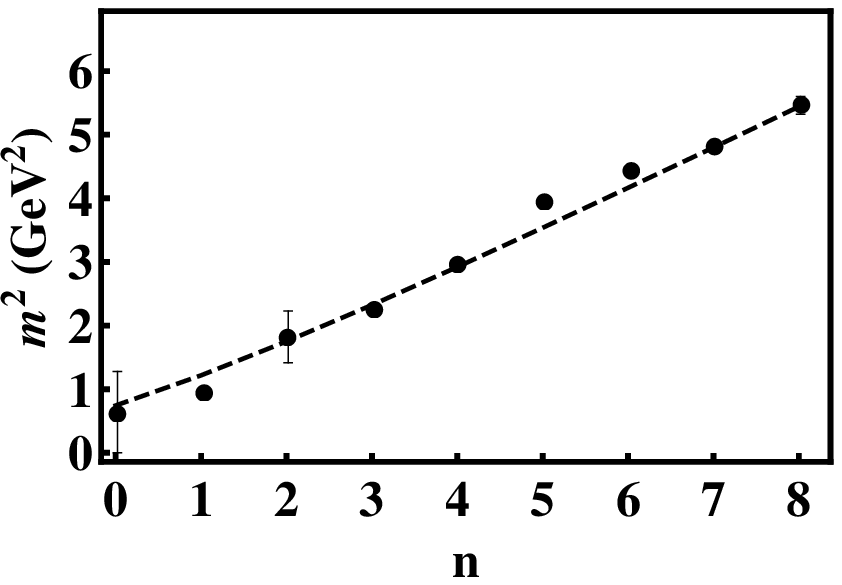,width=4.3cm,height=3.8cm} \epsfig{figure=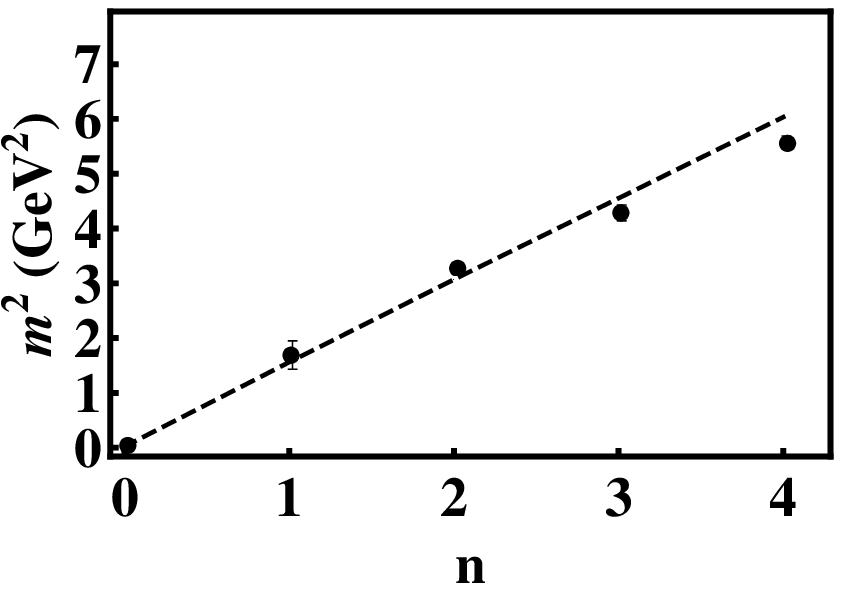,width=4.3cm,height=3.8cm}} \caption{Regge trajectory for $f_0$
(left panel) and pion (right panel) from  the Dynamical AdS/QCD model with $\Lambda _{\text{QCD}}=0.3$ GeV (dashed line). Experimental data from
\cite{pdg}. } \label{Fig1}
\end{figure}

 In reference \cite{dePaulaPRD09}, within the context of
higher spin mesonic states, we show that the metric
\begin{equation}
A(z)= Log(z \Lambda_{QCD}) + \frac{1+\sqrt{3}}{2S+\sqrt{3}-1}\frac{(z\Lambda _{\text{QCD}})^{2}}{%
1+e^{(1-z\Lambda _{\text{QCD}})}}  \label{cnew}
\end{equation}%
gives
\begin{equation}
\mathcal{V}_{S}\left( z\right) \overset{z\rightarrow \infty }{%
\longrightarrow } \Lambda_{QCD}^4\left(1+\sqrt{3}\right)^{2}\;z^{2}\ ,
\label{eqn:effP}\end{equation}%
as the leading infrared contribution to the effective potential. This leads to a satisfactory description of the meson mass spectrum with nearly
universal Regge slopes, just using the natural scale of $\Lambda_{QCD}$, without any further tuning of parameters. (The spin dependent factor in
Eq. (\ref{cnew}) is required by universality of effective IR potential (\ref{eqn:effP})). A good analytical approximation to the spectrum
calculated with $\Lambda _{\text{QCD}}=0.3$ GeV for $S\ge 1$ and $n \geq 0$ is  $m^2\simeq \frac{1}{10}(11n+9S+2)$, that provides an overall fit
to the radial and high-spin excitations \cite{dePaulaPRD09}.

In this letter we propose a generalization of the Dynamical AdS/QCD model for scalar mesons based on the universality of the form of the effective
potential in the IR limit. We implement a scale transformation in the holographic distance tailored to fit the slope of the scalar Regge
trajectory. We assume the same universal form of the metric as given by
\begin{equation}
A(z)= Log(z \Lambda_{QCD}) + \frac{(\xi z\Lambda _{\text{QCD}})^{2}}{%
1+e^{(1-\xi z\Lambda _{\text{QCD}})}}  \label{cscalar}
\end{equation}%
with $\xi=0.58$  from the fit (see figure 1).  The slope of the Regge trajectory is decreased for the scalar excitations (see figure 1) in respect
to $S=1$ as $\xi<1$. In our model, the size of $f_0(600)$ should be larger than the size of other light mesons. Therefore, $f_0(600)$ comes as a
broad resonance in pionic channels owing to a large overlap between the corresponding amplitudes. More on that will come on what follows (see
figure 2 and Table I).

\begin{figure}[tbh]
\centerline{\epsfig{figure=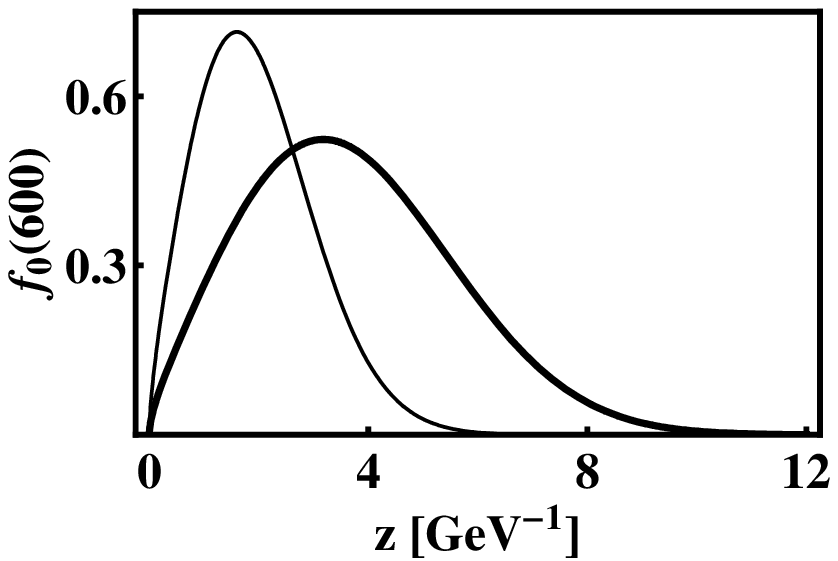,width=4.5cm,height=3cm} \epsfig{figure=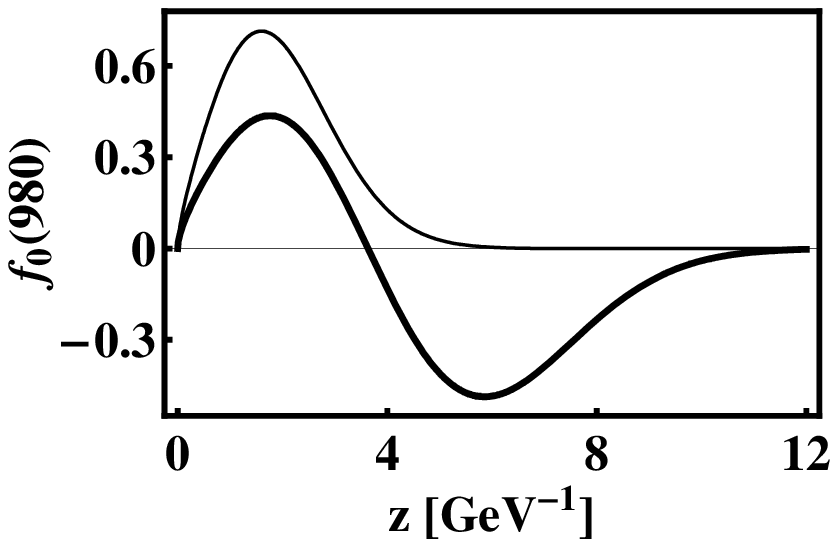,width=4.5cm,height=3cm}}
\centerline{\epsfig{figure=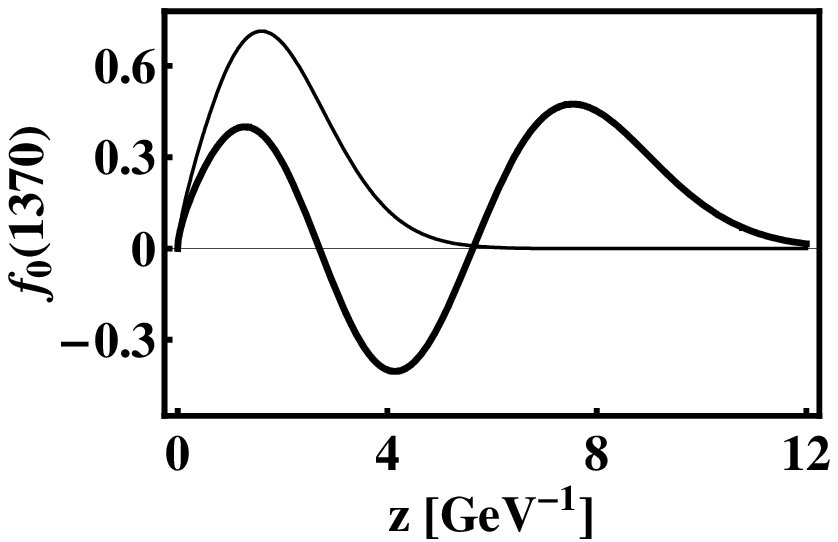,width=4.5cm,height=3cm} \epsfig{figure=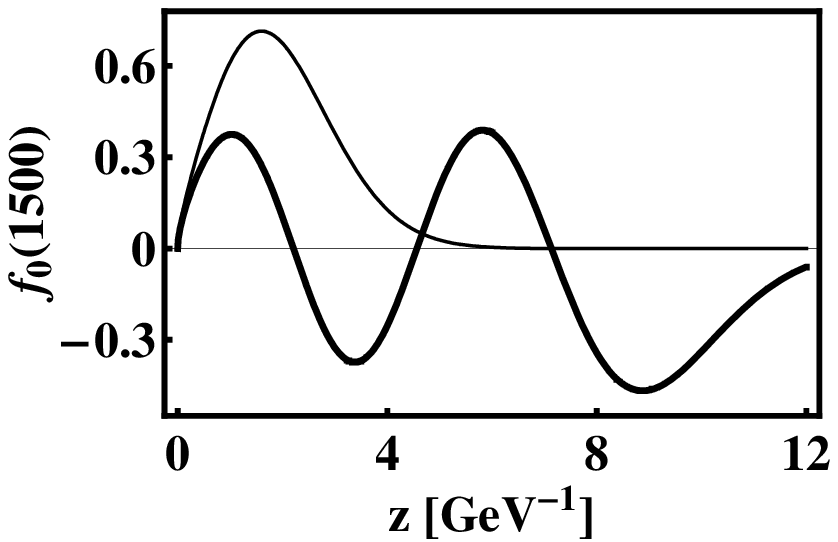,width=4.5cm,height=3cm}} \caption{ Normalized wave
functions for $f_0$ mesons (thick solid line) and pion (solid line) from the Dynamical AdS/QCD model with $\Lambda _{\text{QCD}}=0.3$ GeV.}
\label{Fig2}
\end{figure}

Solving the Sturm-Liouville equation (\ref{sleq}) we obtain the scalar meson spectrum and wave functions. We used the lowest conformal dimension
operator corresponding to a scalar with twist 2 \cite{BrodskyPRL09}.  The twist dimension selects the minimal partonic content of a hadron that is
attributed to its valence wave function. For operators which have the same twist dimension the model at this stage gives in principle a degenerate
spectrum. In particular this difficulty appears in the scalar and pseudoscalar description as we are going to discuss further. We remark that in
our model the scalar state corresponds to a $q\bar q$ meson and it is not identified with a dilaton fluctuation or to a glueball state that has a
different twist dimension.

The trajectories found were in agreement of experimental data of $f_{0}$ family as shown in figure 1. In this picture the sigma meson is the
fundamental state and the other $f_{0}$'s are radial excitation of the sigma. We have included in the plot all $f_{0}$'s present in the Particle
Listing of PDG \cite{pdg}. That agreement may be fortuitous and to support our interpretation we test the physics brought by the structure of
these mesons measured by the $S\to PP$ partial width, that reflects the overlap between the dual string amplitudes of the corresponding mesons.
Therefore, the model should include a description of the string dual to the pion state, as well.

The first striking point is the slope of about 1 GeV$^2$  for the pion Regge trajectory, with a value twice of the one found for the scalars. This
indicates that the scaling factor of the holographic coordinate for the pseudoscalars should be changed in respect to the $f_0$ family.  A scaling
factor of $\xi=0.76$ makes the IR effective potential of the pion the same as the one found the higher spin mesons\cite{dePaulaPRD09}. By allowing
a fine-tuning variation of about 15\% to fit the actual data, we found $\xi=0.88$.

The almost vanishing pion mass is implemented by rescaling the fifth dimensional mass according to $M_{5}^{2} \rightarrow M_{5}^{2}-\lambda z^{2}$
(see \cite{FBF}), where $\lambda$ is uniquely determined as $\lambda=2.19$GeV$^2$. The parameter $\lambda$ gives the strength of the pion coupling
to a scalar background. It will be introduced later as the normalization scale of the scalar decay amplitude. The size of the pion wave function
is considerably smaller than the size of sigma, as shown in figure 2. This reflects the larger slope seen for the pseudoscalar Regge trajectory in
comparison to the scalar one.

The rescale of the string mass can be interpreted as an indication of the coupling between the given mode with those corresponding to higher twist
operators. In a couple channel model this could decrease the mass of the ground state as the parameter $\lambda$ does.

We emphasize that no new free parameter is really associated with the pion Regge trajectory, beyond the almost vanishing pion mass and the
universality of the IR potential as dictated by the observed spectra of light-flavored mesons (apart from the $f_0$ family).

The $f_0$'s partial decay width into $\pi\pi$ are calculated from the overlap integral of the normalized string amplitudes (Sturm-Liouville form)
in the holographic coordinate  dual to the scalars $(\psi_n)$ and pion $(\psi_\pi)$ states,
\begin{equation}
h_{n} =\lambda\,\Lambda_{QCD}^{-\frac32}\int^\infty_0 dz ~\psi_\pi^2(z)\psi_n(z) \ , \label{dwover}
\end{equation}
 We have introduced the parameter $\lambda$ in the transition
amplitude considering that it gives the natural scale for the coupling between the pion and a scalar, as has been obtained through the pion mass
shift. By dimensional analysis one has to consider, that the coupling has dimension of $\sqrt{mass}$ and therefore $\Lambda_{QCD}$ comes to bring
it to the correct dimension. We find that $\lambda\, \Lambda_{QCD}^{-\frac32}= 13~[$GeV$]^\frac12$, for $\Lambda_{QCD}=$ 0.3 GeV,  giving the
results shown in table I.

\begin{table}[htb]
\caption{Two-pion decay width and masses for the $f_{0}$ family. Experimental values from PDG\cite{pdg}. ($^\dagger$Mixing angle of $20^o$.)}
\centering
\begin{tabular}{c c c c c}
\hline\hline Meson & $M_{exp}$(GeV) & $M_{th}$(GeV) &
$\Gamma^{exp}_{\pi\pi}$(MeV) & $\Gamma^{th}_{\pi\pi}$(MeV) \\
[0.5ex] \hline
$f_{0}(600) $ & 0.4 - 1.2    & 0.86 & 600 - 1000 & 602 \\
$f_{0}(980) $ & 0.98 $\pm$ 0.01  & 1.10   & $\sim$ 15 - 80  & 47$^\dagger$  \\
$f_{0}(1370)$ & 1.2 - 1.5    & 1.32   & $\sim$ 41 - 141 & 159   \\
$f_{0}(1500)$ & 1.505 $\pm$ 0.006  & 1.52   & 38 $\pm$ 3  & 42    \\
$f_{0}(1710)$ & 1.720 $\pm$ 0.006 & 1.70   & $\sim$ 0 - 6    & 6  \\
$f_{0}(2020)$ & 1.992 $\pm$ 0.016  & 1.88   &  ---     & 0.0     \\
$f_{0}(2100)$ & 2.103 $\pm$ 0.008  & 2.04   &  ---     & 1.4  \\
$f_{0}(2200)$ & 2.189 $\pm$ 0.013  & 2.19   & ---     &  2.8   \\
$f_{0}(2330)$ & 2.29 - 2.35 & 2.33   &  ---     & 3.2  \\
\hline
\end{tabular}
\label{table:width}
\end{table}

The overlap integral is the dual representation of the transition amplitude $S\to PP$ and therefore the decay width is given by $\Gamma_{\pi
\pi}^{n} = \frac{1}{8\pi}|h_{n}|^{2}\frac{p_\pi}{m_n^2}\ ,$ where $p_\pi$ is the pion momentum in the meson rest frame. The Sturm-Liouville
amplitudes of the scalar (pseudoscalar) modes are normalized just as a bound state wave function in quantum mechanics
\cite{RadyushkinPRD2007,BrodskyPRD2008025}, which also corresponds to a normalization of the string amplitude
\begin{equation}
\int_{0}^{\infty}dz\psi_{m}(z)\psi_{n}(z)=\int_{0}^{\infty}dz \varphi_{m}(z)e^{-(\Phi+3A)}\varphi_{n}(z)=\delta_{mn} \ . \label{norm}
\end{equation}

The overlap integral for the transition amplitude (\ref{dwover}) is naturally damped by increasing the $f_0$ excitation as the destructive
interference comes into scene within the holographic view. A qualitative understanding of that effect can be seen in figure 2, by observing that
within the range of the pion wave function nodes of the higher scalar excitation takes place.

The two-pion partial decay widths for the $f_0$'s present in the particle listing of PDG, are calculated with Eq. (\ref{dwover}) and shown in
Table I. In particular for $f_0(600)$ the model gives a width of about 600 MeV, while its mass is 860 MeV. The range of experimental values quoted
in PDG for the sigma mass and width are quite large as depicted in Table I.  A recent analysis of the sigma pole in the $\pi\pi$ scattering
amplitude from ref.\cite{Caprini06} gives $m_\sigma=$ 441$^{+16}_{-8}$ MeV and $\Gamma_\sigma=544^{+18}_{-25}$ MeV, which in comparison to our
results the width seems consistent while the model mass appears somewhat larger. The analysis of the E791 experiment gives $m_\sigma=$
478$^{+24}_{-23}\pm 17$ MeV and $\Gamma_\sigma=324^{+42}_{-40}\pm 21$ MeV \cite{Aitala_sigma}, and the CLEO collaboration \cite{CLEO2002} quotes
$m_\sigma=$ 513$\pm 32$ MeV and $\Gamma_\sigma=335\pm 67$ MeV, both values of the width smaller than our result. Other analysis of the
$\sigma$-pole in the $\pi\pi \to \pi\pi$ scattering amplitude present in the decay of heavy mesons indicates a mass around 500 MeV \cite{bugg06}.
A rescaling of the string mass as seen necessary for the pion case can lower the sigma mass. We just observe that coupling between the pion with
higher twist string duals can be a source for the effective decrease of the string mass as in the pion case. The width is not strongly affect as
the shift in the string mass mainly dislocates  the squared meson mass by a constant. We do not attempt to fine-tuning the model at this stage.

The $f_0(980)$ is  identified with the first excitation of the string model dual to $q\bar q$ state (see Table I). The model mass is shifted to a
value above the experimental one, i.e., 1.1 GeV compared to 0.98 GeV. The shift of about -0.12 GeV can be attributed to a rescaling of the string
mass as in the sigma case. By increasing the excitation of the scalar meson this shift tends to decrease (see $f_0(1500)$ in Table I). The
experimental values of $\Gamma_{\pi\pi}$ for $f_0(980)$ is too small compared to our result. We introduced a mixing angle for $f_0(980)$ of $\pm
20^o$, that corresponds to a composite nature by mixing, e.g., $s\bar{s}$ with light non-strange quarks\cite{Bediaga}. The  mixing angle absolute
value between $\sim 11^{\circ}$ to $26^{\circ}$ fits $\Gamma^{\pi\pi}$ within the experimental range.

The mass of the higher scalar excitations are consistent with the experimental data in Table I and figure 1. The two-pion partial decay widths of
$f_0(1370)$, $f_0(1500)$ and $f_0(1710)$ are in good agreement to experimental range, without any further assumption.

Before concluding, let us discuss the validity of (8) used to calculate the partial
decay width. Normally, this formula is understood as a reasonable approximation
when the width is small. For the zero excitation, the sigma or $f_0(600)$,
the width is quite wide as shown in table I, and in this case the approximation
could be questionable. A weighted average of the decay rate around the
attributed sigma mass would be preferred, as the masses of two-pion states
in the resonant decay channel spreads out around the resonance mass. That
means that the string amplitude dual to the scalar meson should be coupled
to the two-pion state in the continuum. However, even in this more detailed
picture, the string amplitude dual to $f_0(600)$ heals into a region dictated by
the confinement scale in the fifth dimension, that should not be affected by
the coupling to pions. Therefore, within our holographic model, we believe
that the calculation of the width without resorting to a weighted average is a
reasonable approximation even in the case of the wide sigma meson, because
the relevant physics is dominated by the confinement scale.

In summary, we provide the basic framework to study the $f_0$ family given by excitations of the sigma meson as the string duals to $q\bar q$
states. The classification, spectroscopy and decay is guided by an Holographic view of the scalar string modes obtained from a Dynamical AdS/QCD
model\cite{dePaulaPRD09} of coupled dilaton-gravity background solution of the Einstein equations.  The deformation of the anti-de Sitter metric
encodes confinement by the area law behavior of the Wilson loop. Assuming the universality of the metric, apart a scale deformation, for the
scalar,  pseudoscalar and higher spin string modes dual to $q\bar q$ states, the Regge-like spectrum of the radial excitations of $f_{0}$ was
obtained and the slope about 0.5 GeV$^2$ fitted to the data. The almost vanishing pion mass was obtained adding a rescaled string mass. The
decrease of the string mass can also improve the description of the lower excitations of the $f_0$ family, which can be attributed to the coupling
between $q\bar q$ states to more complex configurations.

A nice point about the interpretation of holographic QCD comes from
a recent work of Brodsky and T\'eramond \cite{BrodskyPRL09}, that
related the wave equation in the Sturm-Liouville form to the squared
mass operator eigenvalue equation for the valence component of the
meson light-front wave function. In that respect, the effective
potential that we have derived should incorporate the coupling
between the valence wave function to all other higher
Fock-components of the wave function, indicating a possible new
class of AdS/QCD models with the coupling between twist 2 and higher
twist modes. The experimentally known partial decay width into
$\pi\pi$ give further support to the proposed classification scheme
for $f_0(600)$, $f_0(980)$, $f_0(1370)$, $f_0(1500)$ and
$f_0(1710)$, while in the particular case of $f_0(980)$ it should
mix with more exotic structures as $s\bar s$ with an estimated
mixing angle around $\pm 20^o$.

Our phenomenological study of a 5d dynamical model consistent with
the available experimental data, may constitute an useful guidance
for the construction of 10d supergravity theories, that allows an
effective 5d perspective and incorporates QCD-like properties. A 10d
supergravity theory dual to QCD is a wishful goal, but it is still
undelivered (see \cite{csaki} and \cite{waynebianchi}). Our
bottom-up approach, looks promising from the point of view of the
phenomenology, but still there is a long journey to a 10d theory.

We acknowledge partial support from the Brazilian agencies CAPES, CNPq and FAPESP.


\begin{thebibliography}{99}
\bibitem{pdg} C. Amsler et al. (Particle Data Group), Phys. Lett. B \textbf{667} (2008) 1.
\bibitem{Jaffe} R. Jaffe, Phys. Rev. D \textbf{15} (1977) 28.
\bibitem{Maiani} L. Maiani, F. Piccinini, A. D. Polosa, V. Riquer, Phys. Rev. Lett. \textbf{93}   (2004) 212002;
G. 't Hooft, G. Isidori, L. Maiani, A. D. Polosa, V. Riquer, Phys.
Lett. B \textbf{662} (2008) 424.
\bibitem{ACCIARRI01H} M. Acciarri et al., Phys. Lett. B \textbf{501} (2001) 173.
\bibitem{ABE04} K. Abe et al., Eur. Phys. J. C \textbf{32} (2004) 323.
\bibitem{dePaulaPRD09} W. de Paula, T. Frederico, H. Forkel and M. Beyer, Phys. Rev. D \textbf{79} (2009) 075019; PoS LC2008:046 (2008).
\bibitem{PolchinskiPRL2002} J. Polchinski and M. Strassler, Phys. Rev. Lett. \textbf{88} (2002) 031601.
\bibitem{Karch} A. Karch, E. Katz, D.T. Son and M.A. Stephanov, Phys. Rev. D \textbf{74} (2006) 015005.
\bibitem{BrodskyPRL09} G. F. de T\'{e}ramond and S. J. Brodsky, Phys. Rev. Lett. \textbf{102} (2009)  081601.
\bibitem{Kiritsis} U. G\"{u}rsoy, E. Kiritsis and F. Nitti, JHEP \textbf{0802}  (2008) 019; JHEP \textbf{0802} (2008) 032.
\bibitem{FBF} H. Forkel, M. Beyer and T. Frederico, JHEP \textbf{0707} (2007) 077; Intl. J. Mod. Phys. E \textbf{16}
(2007) 2794.
\bibitem{FK} H. Forkel and E. Klempt, Phys. Lett. B \textbf{679} (2009)
77.
\bibitem{Maldacena_Conjecture} J. Maldacena, Adv. Theor. Math. Phys. \textbf{2} (1998) 231.
\bibitem{Klebanov:2000hb} I.~R.~Klebanov and M.~J.~Strassler, JHEP {\bf
0008} (2000) 052.
\bibitem{Klebanov:2000nc} I.~R.~Klebanov and A.~A.~Tseytlin, Nucl.\ Phys.\  B {\bf 578}
(2000) 123.
\bibitem{Maldacena:2000yy} J.~M.~Maldacena and C.~Nunez, Phys.\ Rev.\ Lett.\  {\bf 86}
(2001) 588.
\bibitem{Berg:2006xy} M.~Berg, M.~Haack and W.~Mueck,  Nucl.\ Phys.\  B {\bf
789} (2008) 1.
\bibitem{Papadopoulos:2000gj} G.~Papadopoulos and A.~A.~Tseytlin, Class.\ Quant.\ Grav.\  {\bf 18} (2001)
1333.
\bibitem{Bediaga} I. Bediaga, F. Navarra and M. Nielsen, Phys. Lett. B \textbf{579}, 59 (2004).
\bibitem{tHooft} G. 't Hooft, Nuclear Physics B\textbf{72}, 461 (1974).
\bibitem{Boschi} H. Boschi, N. Braga and H. Carrion, Eur. Phys. J. \textbf{C 32}, 529 (2004); Phys. Rev. \textbf{D 73}, 047901 (2006);
G. F. de T\'{e}ramond and S.J. Brodsky, Phys. Rev. Lett. \textbf{94}, 0201601 (2005);
\bibitem{Afonin:2009pd} S.~S.~Afonin, Phys.\ Lett.\  B {\bf 675}, 54 (2009);
S.~S.~Afonin, Phys.\ Lett.\  B {\bf 678}, 477 (2009)
\bibitem{Schmidt_Scalar} A. Vega and I. Schmidt, Phys. Rev. D
\textbf{78}, 017703 (2008).
\bibitem{Colangelo_Scalar}P.~Colangelo, F.~De Fazio, F.~Jugeau and S.~Nicotri, Phys.\ Lett.\  B {\bf 652} (2007) 73;
P. Colangelo, F. De Fazio, F. Giannuzzi, F. Jugeau, and S. Nicotri, Phys. Rev. D \textbf{78}, 055009 (2008).
\bibitem{kelley} T. Gherghetta, J. I. Kapusta and T. M. Kelley,
Phys. Rev. D \textbf{79}, 076003 (2009).
\bibitem{Witten98} E. Witten, Adv. Theor. Math. Phys. \textbf{2}, 505
(1998).
\bibitem{RadyushkinPRD2007} H. Grigoryan and A. Radyushkin, Phys. Rev. D \textbf{76}, 095007 (2007).
\bibitem{BrodskyPRD2008025} S. J. Brodsky and G. F. de T\'{e}ramond, Phys. Rev. D \textbf{78}, 025032 (2008).
\bibitem{Caprini06} I. Caprini, G. Colangelo and H. Leutwyler, Phys. Rev. Lett. \textbf{96}, 132001 (2006).
\bibitem{Aitala_sigma} E. M. Aitala et al. (Fermilab E791 collaboration), Phys. Rev. Lett. \textbf{86}, 770 (2001).
\bibitem{CLEO2002} H. Muramatsu et al. (CLEO collaboration), Phys. Rev. Lett. \textbf{89}, 251802 (2002).
\bibitem{bugg06} D. V. Bugg, Eur. Phys. Jour. C \textbf{47}, 57 (2006); AIP Conf. Proc. {\bf 1030}, 3 (2008).
\bibitem{csaki} C. Csaki and M. Reece, JHEP \textbf{05} (2007) 062.
\bibitem{waynebianchi} M. Bianchi and W. de Paula, JHEP \textbf{1004} (2010) 113.
\end{thebibliography}
\end{document}